# Waltzing of a Helium Pair in Tungsten: Migration Barrier and Trajectory Revealed from First-Principles

J. G. Niu[1,2], Q. Zhan[1a], and W. T. Geng[1,3 b]

[1] School of Materials Science & Engineering, University of Science and Technology Beijing, Beijing 100083, China

[2] Department of Mechanical Engineering, Hebei University, Baoding 071000, China

[3] Psi Quantum Materials LLC, Laiwu 271100, China

**Abstract**

Despite well documented first-principles theoretical determination of the low migration energy (0.06 eV) of a single He in tungsten, fully quantum mechanical calculations on the migration of a He pair still present a challenge due to the complexity of its trajectory. By identifying the six most stable configurations of the He pair in W and decomposing its motion into rotational, translational, and rotational-translational routines, we are able to determine its migration barrier and trajectory. Our density functional theory calculations demonstrate a He pair has three modes of motion: a close or open circular two-dimensional motion in (100) plane with an energy barrier of 0.30 eV, a snaking motion along [001] direction with a barrier of 0.30 eV, and a twisted-ladder motion along [010] direction with the two He swinging in the plane (100) and a barrier of 0.31 eV. The graceful associative movements of a He pair are related to the chemical-bonding-like He-He interaction being much stronger than its migration barrier in W. The excellent agreement with available experimental measurements (0.24-0.32 eV) on He migration makes our first-principles result a solid input to obtain accurate He-W interatomic potentials in molecular dynamics simulations.

[a] E-mail: qzhan@mater.ustb.edu.cn

[b] E-mail: geng@ustb.edu.cn

## 1. Introduction

With high melting point, high thermal conductivity as well as low sputtering yield for light elements, tungsten is considered to be a promising candidate for the first-wall and divertor plate in fusion reactors.[1] In working conditions, tungsten is subject to irradiation of H isotopes, He, and neutron, which leads to bubble formation, surface blistering, and hence embrittlement. Understanding of He induced embrittlement of W requires good knowledge of all basic processes controlling microstructural evolution, including diffusion and accumulation of He atoms in the host.[2-4] Strong attraction between interstitial He atoms in W makes it easy to form He pairs or small clusters even in the absence of vacancies or other crystalline defects.[5] To learn how to prevent, or, at least slow down He bubble growth in W, it is indispensable to gain knowledge of the energy barrier and trajectory of He migration in W.

There is, unfortunately, a discrepancy on the migration barrier of a single interstitial He in W, among experimental measurements, density functional theory (DFT) calculations, and molecular dynamics simulations. Using DFT method, Becquart and Domain [5] calculated this value to be 0.06 eV, much smaller than experiment, which ranges from 0.24 to 0.32 eV [6-7]. Xiao et al. have carefully examined the possible numerical error introduced by pseudopotential method and performed all-electron calculations to make comparison,[8] exactly the same energy barrier was obtained. Since a He-He pair in W has a strong binding (0.98 eV/pair) [5], it was suggested that what was measured was not the migration of a single He, but rather a He pair or even a small He cluster instead. On the other hand, molecular dynamics simulations using various interatomic potentials have also been performed on the diffusion of the He clusters in W.[9] A number of counter-intuitive results were reported such that $He_3$ and $He_4$ clusters migrate slower than a single He, while $He_2$ and $He_5$ clusters are faster than a single He. Nevertheless, the large deviation of the calculated migration energy of a single He (0.13 eV) from DFT result (0.06 eV) cast some serious doubt on the accuracy of molecular dynamics determination.

Here we attempt to tackle the problem of migration of a He pair in W in the fully quantum mechanical level using first principles DFT method. The trajectory of a He pair moving in a solid can be expected to be much more complex than that of a single He atom due to the much increased degrees of freedom. The approach we have employed has two key steps. Firstly, we need to identify the most stable configurations of a He pair in W, which are presumably the local energy minima on the migration path. Secondly, we need to decompose the motion of a He pair into rotational, translational, and rotational-translational routines. By connecting these routines on the probable path, we then should be able to determine migration barrier and trajectory of a He pair.

**2. Computational Details**

The first-principles DFT calculations were carried out using the Vienna *ab initio* simulation package (VASP).[10] The electron-ion interaction is described using the projector augmented wave method [11] and the exchange correlation potential using the generalized gradient approximation (GGA) in the Perdew-Burke-Ernzerhof form [12]. The energy cutoff for the plane wave basis set was set to 480 eV for all calculated systems to assure a good description of the closed shell of He 1$s$ electrons. A 4×4×4 128-atom body-centered-cubic (*bcc*) W supercell was employed in this work, which has been shown to be adequate in literature, and the lattice constant for *bcc* W was calculated as 3.17 Å. The Brillouin zone integration was performed within the Monkhorst-Pack scheme, using 2×2×2 mesh for the 128-atom W supercell. Test calculations on a number of selected configurations using a 3×3×3 mesh shows that the numerical error on relative energy caused by a 2×2×2 mesh is within 0.01 eV. Nudged elastic band (NEB) method was adopted to calculate the energy.[13] Five images were linearly interpolated between each pair of neighboring local energy minima on the migration track. In the NEB calculations, the volume of the supercell was fixed and the atomic positions were relaxed.

## 3. Stable Configurations of a He pair

The most stable interstitial site to accommodate He is known to be the tetrahedral interstitial site (TIS).[5] Becquart and Domain[5] investigated eight configurations for a TIS He pair with varying separation distance, and identified the most stable structure. We note, however, although TIS is preferable to the octahedral interstitial site (OIS) for a single He, OIS-TIS or OIS-OIS combinations for a He pair still need to be examined in search of the low-lying states, some of which serve as the intermediate state between neighboring most stable positions in the matrix on the way of migration. In Fig. 1(a), we signify the ten configurations (*TA* to *TJ*, circles in red), eight of which have been examined in Ref. [5], together with four OIS-OIS combinations (*OW* to *OZ*, circles in green). As for TIS-OIS combination, we calculated only the *OA* configuration. The optimized structures of the six most stable pair configurations are plotted in Fig.1 (b). In Table 1, we list the calculated relative energies of different He pair configurations described in Fig. 1(a). Among them all, the most stable one is configuration *E*, in consist with the conclusion reached in Ref. [5]. Therefore, we have chosen *E* as the initial and final position (separated by a lattice constant, *a*) when tracing the migration of the He pair.

**Fig. 1. (a) Low energy positions of a He-He pair in *bcc* W. Red circles represent tetrahedral interstitial sites (TIS) and green ones are octahedral interstitial sites (OIS). For TIS-TIS pairs, the first He is in site *T*; and for OIS-OIS pairs, the first He is in site *O*. (b) Optimized geometry of *TA*, *TB*, *TC*, *TE*, *TO*, and *OW*, denoted by *A*, *B*, *C*, *E*, *TO*, and *W*, respectively.**

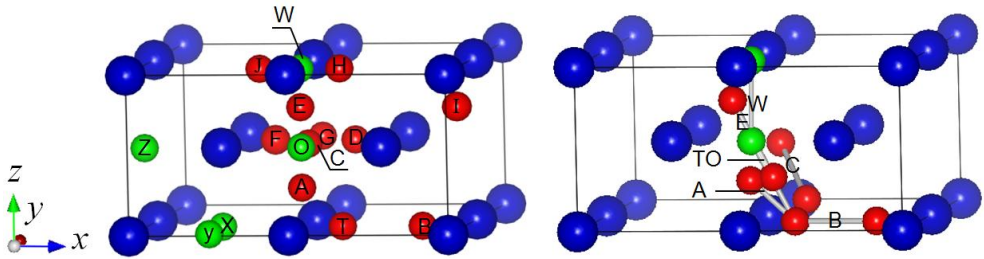

**Table 1. Relative energies of He pair configurations in *bcc* W, as depicted in Figure 1 (a). Both initial and optimized final He-He distances are listed.**

| Configuration | Initial $d_{\text{He-He}}$ (Å) | Final $d_{\text{He-He}}$ (Å) | Relative energy (eV) |
|---|---|---|---|
| A | $a\sqrt{2}/4 = 1.12$ | 1.44 | 0.28 |
| B | $a/2 = 1.59$ | 1.50 | 0.06 |
| C | $a\sqrt{6}/4 = 1.94$ | 1.53 | 0.02 |
| D | $a\sqrt{2}/2 = 2.24$ | 1.65 | 0.58 |
| E | $a\sqrt{10}/4 = 2.51$ | 1.51 | 0 |
| F | $a\sqrt{3}/2 = 2.75$ | $\to C$ | - |
| G | $a\sqrt{14}/4 = 2.97$ | $\to E$ | - |
| H | $a = 3.17$ | 3.20 | 0.98 |
| I | $a\sqrt{18}/4 = 3.36$ | 3.45 | 1.15 |
| J | $a\sqrt{5}/2 = 3.54$ | 3.40 | 0.90 |
| TO | $a\sqrt{5}/4 = 1.77$ | 1.48 | 0.20 |
| W | $a/2 = 1.59$ | 1.50 | 0.44 |
| X | $a\sqrt{2}/2 = 2.24$ | $\to A$ | - |
| Y | $a\sqrt{3}/2 = 2.75$ | 1.56 | 0.59 |
| Z | $a = 3.17$ | 3.31 | 2.34 |

## 4. Elemental Routines of Migration Steps

### A. Translation of a He pair

Although we can anticipate simply from intuition that a rigid translation of a He-He pair, in which the two He are about 1.5 Å apart, in crystalline W would encounter significant energy barrier for there are W atoms in all the direction they are moving and therefore a He pair will most probably keep rotating during migration, it is still advisable to have the quantitative knowledge of how difficult they will be. We have examined three translational motions, namely, those along *x*, *y*, and *z* axis respectively (see Fig. 2). We note that in these motions, the He-He separation has been allowed to change in response to the squeeze by adjacent W atoms. Apparently, when moving along *x*, the He pair will be cut into two separated atoms. The energy needed, according to our calculation, is 1.49 eV, larger than the binding energy of a He pair. It indicates a demolition of the pair. To move in *y* direction, the two He atoms will have to squeeze through a pair of W atoms which are *a* (3.17 Å) apart. The He-He distance will be reduced significantly to 1.30 Å, and the energy needed is also as high as 1.86 eV. The translational motion along z axis, however, is quite different. The two He do not need to pass by their respective adjacent W simultaneously in this case, so they have the space to adjust their separation to avoid severe compression. The calculated energy barrier for this motion is only 0.44 eV, much lower than in *x* and *y* directions.

**Fig. 2. Translational motion of a He pair in *bcc* W. The He atoms in the starting position (configuration *E*) are in red, and those passing through the barriers in three directions are in green. Note that the pair is broken when moving along *x*, while seriously compressed along *y* direction.**

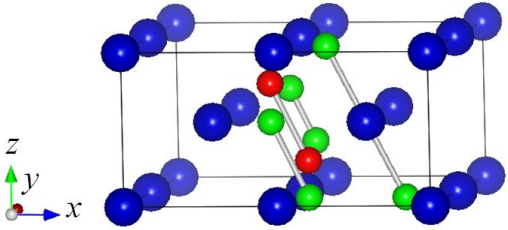

**B. Rotation of a He pair**

A pure rotation of a He pair in the most stable configuration *E* keeps its center of mass in an OIS site. The final state of a pure rotation can only be a geometry equivalent to *E*. It is rather straightforward to figure out that there exist only two types of rotation. One rotation has a trajectory only in *x-z* plane. The configuration *B*, which is 0.06 eV higher in energy than *E*, is right in the middle of this movement (Fig. 3(a)). The angle of rotation is $\theta = 44°$. Interestingly, the energy barrier for this rotation is exactly the same as that for a single interstitial He in W. We denote this rotation routine as $E \rightarrow B \rightarrow E$. The other rotation (Fig. 3(b)) is three dimensional, from the *x-z* plane to the *x-y* plane with $\theta = \varphi = 68°$. The energy barrier for this process is calculated to be 0.17 eV, and we name this routine as $E \rightarrow V \rightarrow E$.

**Fig. 3. The calculated Energy barriers for a two-dimensional (a) and a three-dimensional rotation (b) of a He pair in configuration *E*. The local environments of He pair in the initial, peak, and final positions are shown.**

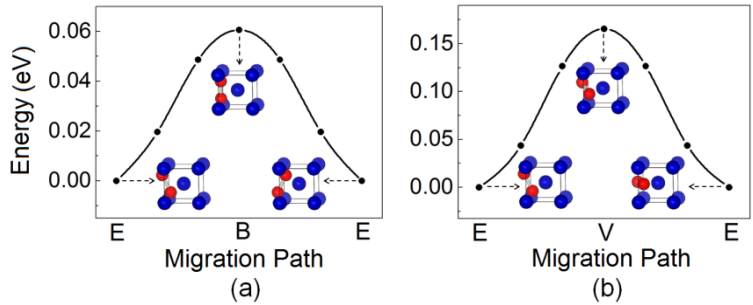

C. Rotational-Translational routines

Inspired by the discovery that both rotational barriers are quite low (0.06 & 0.17 eV) and the migration barrier for rigid movement of the He pair along the <100> which is roughly parallel to the He-He axis, 0.44 eV, is only moderately higher than the diffusion barrier measured for He in experiment, 0.24-0.32 eV, [6-7] we were strongly motivated to search combined rotational-translational routines with low energy barriers. To avoid unnecessary notation complication, we have always made the He pair starting from configuration *E* as displayed in Fig.1 (b), Fig. 2, and Fig. 3.

**Fig. 4. The calculated Energy barriers of three types of rotational-translational routines for a He pair in configuration *E*. The local environments of He pair in the initial, peak, and final positions are shown.**

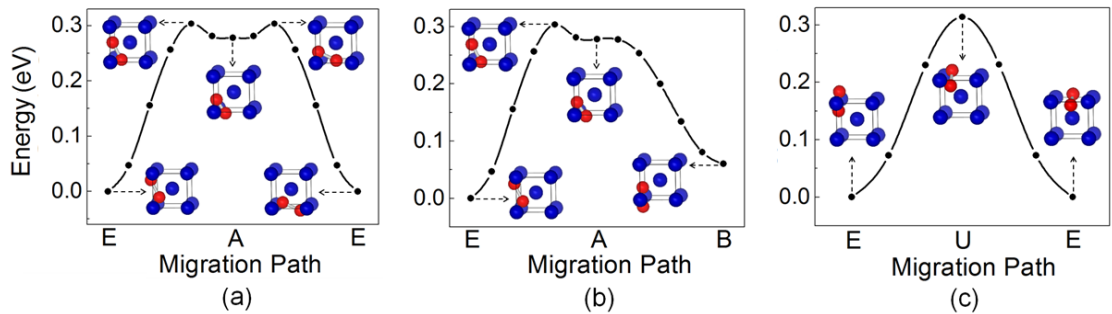

First, let us study the motion from configuration *E* to *A* (see Fig. 1(b)), a two-dimensional movement. On the way moving along *z* direction, we now allow the He atoms to adjust both their separation and center of mass. It can be seen clearly that the peak of the barrier should be around configuration *TO*. If the He pair is heading rightward, that is, along *x*, it will end up with an equivalent *E* which is roughly parallel to *x*. We will denote this routine as *E*→*A*, which apparently has a reverse process of *A*→*E*. If however, the He pair chooses to move downward, it will need to rotate clock-wisely at *A* and reach the configuration *B*. Our DFT-NEB calculations demonstrate that there is no barrier in the routine *A*→*B*. In Fig. 4, we plot the migration paths *E*→*A*→*E* (panel a) and *E*→*A*→*B* (panel b), both of which has an energy barrier of 0.30 eV. If the pair wants to migrate along *y* axis to reach the nearest equivalent *E*, it has to rotate in the *x-z* plane simultaneously, in an attempt to avoid been squeezed by the two adjacent W atoms. The transition state in this process, shown in Fig. 4(c) and denoted as *U*, is not very close in position to any of the configurations displayed in Fig. 1. The calculated energy barrier of this routine, here denoted as *E*→*U*→*E*, is 0.31 eV. It has to be pointed out that the extreme similarity in energy barrier between *E*→*A*→*E* and *E*→*U*→*E* might prompt us to wonder if this small difference is simply a numerical error, but the two should be exactly the same instead. We argue that this is mainly a coincidence. In a different metal, Fe for instance, the two barriers could remarkably differ from each other. This is certainly an interesting issue to study in future works.

## 5. Migration Steps

Having uncovered the basic movement routines of a He pair, we can now construct *full* migration steps. By *full* we mean the center of mass of a He pair move from position $\vec{P}$ (an OIS) to $\vec{P} + \vec{a}$, where is a unit vector defining <100> with a length of lattice constant (3.17 Å for *bcc* W).

## A. Circular motion in (100)

It is easy to perceive from Fig. 4(a) that if we continue the rotational-translational routine $E \rightarrow A \rightarrow E$ with rotation $E \rightarrow B \rightarrow E$, then repeat this process rightward, we will end up with a full migration step of a He pair along $x$ direction. By this step, the He pair circumvents the W atom which is exactly in its way rightward. The trajectory of this movement forms a lower half of one circle. We note that this displacement can be done equally if the pair chooses the upper half of this circle. In panel (a) of Fig. 5, we draw the trajectory of this migration step, where two unit bcc cells are shown to give a better perception. To guide the eye, we make the head He red and the tail He green, and number sequentially the pair in different positions. Also, only some intermediate pairs are numbered, bonded, and have the normal size, other points are displayed using small circles. The same method is also used in panels (b) and (c) for the other two kinds of motions.

**Fig.5. Trajectories of three kinds of migration steps for a He pair in bcc W. To guide the eye, we make the head He red and the tail He green, and number sequentially the pair in different positions. Also, only some intermediate pairs are numbered, bonded, and have the normal size, other points are displayed using small circles. (a) Circular motion in (100); (b) Snaking motion along [001]; (c) Twisted ladder motion along [010].**

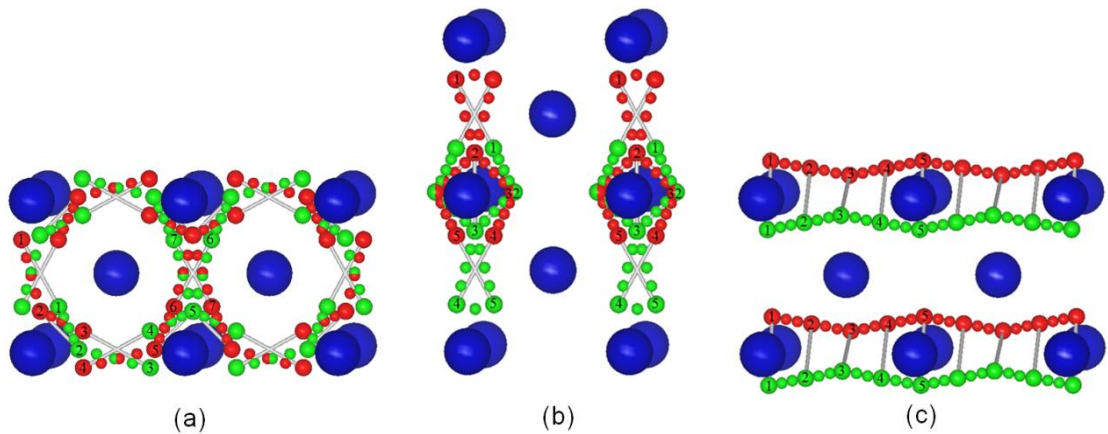

## B. Snaking motion along [001]

On the other hand, if we combine the downward ($z$ direction) rotational-translational routine $E \rightarrow A \rightarrow B$ with its reverse $B \rightarrow A \rightarrow E$, and then supplement it with a pure rotation $E \rightarrow B \rightarrow E$, then we can realize a full downward migration step. The trajectory of this step shows a snaking movement, as is clearly seen in panel (b) of Fig. 5. We note that similar to the circular motion, at each $E$ configuration, the He pair can easily change slightly the orientation of its axis via a pure rotation $E \rightarrow B \rightarrow E$, hence the reflection symmetry of the trajectory.

## C. Twisted-ladder motion along [010]

Finally, for the migration step of the He pair along $y$ direction, we need the $E \rightarrow U \rightarrow E$ routine. We find that with two consecutive $E \rightarrow U \rightarrow E$, a He pair in configuration E moves from position $\vec{P}$ (an OIS) to $\vec{P} + \vec{a}$, hence the achievement of a full migration step. In this step of, the center of mass of the He pair moves in a straight line, with the two He atoms swinging in the (100) plane. The trajectory of the swinging He-He bonds looks very much like a twisted ladder, as is seen in Fig. 5(c).

We have to emphasize that because of the low barrier (0.17 eV) of pure rotation $E \rightarrow V \rightarrow E$, through which the pair can change its axis orientation, a He pair can make easy transition among all three migration trajectories at any $E$ configuration. Therefore, the migration of a He pair initially formed with any given orientation would be three-dimensional in practice.

Finally, we want to mention the transformation between configurations $E$ and $C$. The former is only 0.02 eV more stable than the latter. This means that upon formation, configuration $C$ has nearly the same probability to appear just as $E$, especially at elevated temperatures. We have calculated the energy barrier for an $E \rightarrow C$ transformation and find the barrier is as low as 0.06 eV, suggesting an easy interchange between the configurations.

## 6. Chemical bonding-like He-He interaction

In an attempt to elucidate the underlying force responsible for the associative migration of interstitial He pairs in W, we have investigated the interaction between the two He against their interatomic distance for a configurations *A*, *B*, *E*, and *U*, which are relevant in their migration steps. Each configuration of a He-He pair defines a straight line and we study an ideal movement of the two He on this line. It is worth noting that the two atoms will very likely not move exactly along the pair axis, but near each local energy extreme, the trajectory of each atom is not expected to deviate much from that axis. We display in Fig. 6 the calculated formation energy of a He pair in bcc W with a varying interatomic distance, in reference to perfect W and free standing He atom. For each He-He separation, we have fixed the positions of the two He and allowed all the other W atoms to relax upon energy minimization.

**Fig.6. Formation energy of a He pair in bcc W with a varying interatomic distance, in reference to perfect W and free standing He atom.** *A*, *B*, *E*, and *U* **are distinct configurations of a He pair which define the migration path (See Fig. 4).**

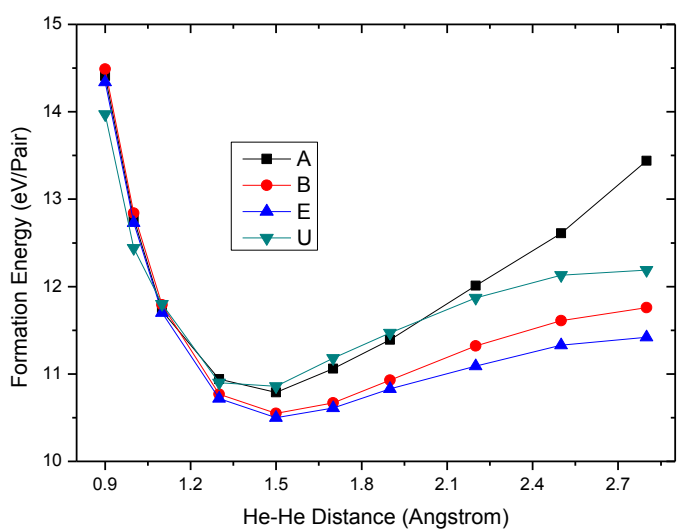

The variation of the formation energy can be simply taken as the interaction of the two He atoms. Very interestingly, we find that the interaction between interstitial He in bcc W is quite similar to a typical chemical bond, except for the long distance end, where periodicity of the W crystal takes effect. In each direction, the He-He attraction is at least as strong as 1 eV, strong enough to hold the pair when passing the 0.30-0.31 eV barriers. More important is that such a powerful attraction is not very short-ranged. This feature is also indispensible in holding the pair in migration.

## 7. Concluding remarks

To summarize, we have identified, using density functional theory calculations, the six most stable configurations of the He pair in W and decomposed its motion into rotational, translational, and rotational-translational routines whose energy barrier can be separately determined. With these basic routines, we are able to construct the full migration steps of a He pair. It is revealed that a He pair has three modes of motion: a close or open circular two-dimensional motion along He-He axis in (100) plane with an energy barrier of 0.30 eV, a snaking motion along [001] direction with a barrier of 0.30 eV, and a twisted-ladder motion along [010] direction with the two He swinging in plane (100) and a barrier of 0.31 eV. The three modes of migration are interchangeable at any allowed position of the most stable configuration and the averaged barrier is estimated to be 0.30 eV.

The calculated energy barrier for a He pair in W, 0.30 eV, is in very good agreement with available experimental measurements (0.24-0.32 eV) on He migration, which we speculate is for He pairs rather than individual He atoms. The graceful movements of a He pair are possible because the chemical-bonding-like He-He interaction is greatly stronger than its migration barrier in W. It is noteworthy that molecular dynamics simulations [10] suggested a He pair could migrate faster than a single He atom, in sharp contrast to our first-principles result. This is a strong indication that the DFT based first-principles results reported here should serve as a solid input to obtain accurate He-W interatomic potentials in molecular dynamics simulations.

The idea to decompose the motion of a He pair into rotational, translational, and rotational-translational routines, and then construct full migration steps by combination of such routines can be readily applied to other metals or alloys with the same (*bcc*) or different crystal structures. Identification of the three modes of movement also strongly suggests the associative migration of small He clusters in W is very likely in linear configurations.

**Acknowledgments**

We are grateful for support of the National Magnetic Confinement Fusion Program (Grant Numbers 2011GB108002 and 2014GB104003) of China. J.G.N. thanks Dr. Wei Hao for helpful discussions. The calculations were performed on the Quantum Materials Simulator of USTB.